
\magnification=\magstep1
\tolerance=10000
\baselineskip=22 pt

\centerline{{\bf DYNAMICAL LORENTZ SYMMETRY BREAKING}}
\centerline{{\bf FROM {\it 3+1} AXION-WESS-ZUMINO MODEL}} 
\bigskip

\centerline{{\bf A. A. Andrianov}
\footnote{$^\natural$}{{\tt e-mail:  andrian@snoopy.phys.spbu.ru}}}
\centerline{\sl Department of Theoretical Physics,}
\centerline{\sl Sankt-Petersburg State University,}
\centerline{\sl 198904 Sankt-Petersburg, Russia}
\centerline{ and}
\centerline{\sl  Departament d'Estructura i Constituents de la Mat\`eria}
\centerline{\sl Universitat de Barcelona }
\centerline{\sl  Diagonal, 647, 08028 Barcelona, Spain}

\medskip

\centerline{{\bf R. Soldati}
\footnote{$^\flat$}{{\tt e-mail: soldati@bo.infn.it}}}
\centerline{\sl Dipartimento di Fisica "A. Righi", Universit\'a
di Bologna and Istituto Nazionale}
\centerline{\sl di Fisica Nucleare, Sezione di Bologna,
40126 Bologna, Italia}

\medskip

\centerline{{\bf L. Sorbo}
\footnote{$^\sharp$}{{\tt e-mail: sorbo@sissa.it}}}
\centerline{\sl International School for Advanced Studies}
\centerline{\sl 34013 Trieste, Italia}

\vfill
\noindent 
DFUB-98-12 \hfill June 98   
\eject

\centerline{ABSTRACT}
\vskip 1.5 truecm
\noindent 
{\it We study the renormalizable abelian vector-field models in the presence
of the Wess-Zumino
interaction with the pseudoscalar matter. The renormalizability is achieved
by supplementing the standard kinetic term of vector fields with higher 
derivatives.
The appearance  of fourth power of momentum in the vector-field propagator
leads to the super-renormalizable theory in which the $\beta$-function,
the vector-field renormalization constant and the anomalous mass dimension
are calculated exactly. It is shown that this model has the infrared
stable fixed point and its low-energy limit is non-trivial. 
The modified effective potential for the pseudoscalar matter 
leads to the possible occurrence of dynamical breaking of the Lorentz symmetry.
This phenomenon is related to the modification of Electrodynamics
by means
of the Chern-Simons (CS)
interaction polarized along a constant CS vector. Its presence makes the
vacuum optically active that has been recently estimated from
astrophysical data. We examine two possibilities for the CS vector to
be time-like or
space-like, under the assumption that it originates from v.e.v.
of some pseudoscalar matter and show that only the latter one is consistent
in the framework of the AWZ model, because a time-like CS vector makes 
the vacuum unstable under pairs creation
of tachyonic photon modes with the finite vacuum decay rate.}
\vskip 1.0 truecm
\noindent
PACS numbers : 11.30.Qc, 14.80.Mz

\vskip 1.5 truecm
\centerline{\sl Submitted to Phys. Rev. D}
\vfill\eject

\noindent
{\bf 1.\ Introduction.}
\medskip
The possible occurrence of a very small deviation from the 
Lorentz covariance has 
been considered and discussed some time ago [1], within the context of
the Higgs sector of spontaneously broken gauge theories. There, some
"background" or "cosmological" field is generated, leading to the above
mentioned possible small deviations from the Lorentz covariance, 
within the present experimental limits.

Later on another possibility has been explored to 
obtain a Lorentz- and parity-violating
modification of quantum electrodynamics, by means of addition of a Chern-Simons
lagrangean [2].
Quite recently, Coleman and Glashow [3]
have discussed how Lorentz non-invariant
velocity differences among neutrinos could produce characteristic flavour
oscillations in accelerators and solar neutrino fluxes. 

In all the above investigations, 
the Lorentz symmetry breaking (LSB) has been treated 
phenomenologically by means of some very small but explicit Lorentz 
non-invariant terms which have a clear physical meaning in a privileged
frame. Then, of course, the dynamical (and presumably quantum) origin 
of possible LSB represents 
an interesting problem to be tackled.

One of the possible ways to induce LSB by a dynamical mechanism  
has been recently
argued in the {\it 3+1} dimensional case [4]. Namely,
 the spontaneous breaking of the 
Lorentz symmetry {\it via} the Coleman-Weinberg mechanism [5] has been
revealed for a class of models with the Wess-Zumino interaction between
abelian gauge fields and pseudoscalar axion (AWZ models).

The original motivation for studying  AWZ models was to use them in
resolving the old-standing conflict
between perturbative renormalizability and unitarity, within the context of the 
gauge invariant quantization of {\it 3+1} dimensional abelian gauge models in
the presence of the $U(1)$-chiral anomaly [6].

As a matter of fact, it has been suggested some time ago [7] that gauge 
theories in the presence of chiral
anomalies could be consistently quantized after integration over the gauge
orbits and the introduction of suitable Wess-Zumino fields. Although this idea
has been successfully implemented in low dimensions [8],[9], its application
to the {\it 3+1} dimensional case is still to be achieved, even within the 
standard covariant perturbative approach [6], [10].

For lower-dimensional theories the LSB phenomenon has been 
observed by Hosotani in a series of papers
[11]. He has found that in {\it 2+1} dimensional 
Chern-Simons gauge field theories
coupled to Dirac fermions a spontaneous magnetization arises, leading to the
breaking of $O(2,1)$ symmetry. This remarkable effect might be also related to
the breaking of chiral symmetry, {\it i.e.} to the generation of a dynamical
mass for fermions [12], [13].

In the present paper we continue our exploration of LSB by dynamical 
mechanisms and study it in more details in the renormalizable model for
the abelian
vector-field interacting with the pseudoscalar axion matter.

The Wess-Zumino interaction in this model may be understood as generated
 by quantum effects due to coupling to fermions.
For instance, 
one can start from the above mentioned anomalous gauge model 
with the lagrangean density which describes the
coupling of chiral fermions to an abelian gauge field $A_\mu$,
$$
{\cal L}_0[A_\mu,\psi,\bar\psi]=
\bar\psi\gamma^\mu\left\{i\partial_\mu + e A_\mu P_L\right\}\psi  
- m\bar\psi\psi\ , \eqno(1.1)
$$
where $P_L\equiv (1/2)({\bf 1}-\gamma_5)$. After fermion quantization
it leads to the chiral
anomaly, thereby breaking the classical invariance
under local gauge transformations of the left chiral sector and making 
the serious obstruction to derive a unitary and renormalizable gauge
theory. This barring is essentially induced by the coupling to
longitudinal part of gauge potential, that is in turn  described by
the Wess-Zumino interaction in the following sense.
If we rewrite the gauge potential in eq.~(1.1) as
$$
A_\mu (x) = A_\mu^\bot (x) + A_\mu^\| (x) =
\left(\delta_\mu^\nu -{\partial_\mu\partial^\nu\over \partial^2}\right)A_\nu (x)
+\partial_\mu\chi (x)\ ,
\eqno(1.2)
$$
it is well known that integration over fermion fields drives from the classical
lagrangean density (1.1) - in the limit when the mass $m$ can be disregarded -
to the quantum effective lagrangean
$$
{\cal L}_{{\rm eff}} = {e^3\over 48\pi^2}\chi\tilde F_{\mu\nu}F^{\mu\nu}
+{\cal L}^\bot_{{\rm eff}} [A_\mu^\bot]\ ,
\eqno(1.3)
$$
where the last term indicates the gauge invariant non-anomalous part.
Furthermore, it is also evident that the gauge invariant part
of the effective lagrangean (1.3) is sub-leading within the low momenta 
regime we are here dealing with, as its quadratic part can always be
re-absorbed into renormalization of photon wave function (see below),
whereas the quartic term is of the order $\alpha^2[(k/\mu_{\rm IR})
\ln (k/\mu_{\rm IR})]^4$, 
$k$ and $\mu_{\rm IR}$ being the low energy photon momentum and 
normalization scale
respectively.

When the gauge fields are massless, {\it i.e.} photons, a more realistic
model providing at low energies the Wess-Zumino interaction of (1.3) type is  
QED with the additional Yukawa coupling to a scalar chiral field:
$$
{\cal L}_0[A_\mu,\psi,\bar\psi]=
\bar\psi\gamma^\mu\left\{i\partial_\mu + e A_\mu \right\}\psi  
- m\bar\psi\exp\left(2i\gamma_5 Y \tilde\chi\right)\psi\ , \eqno(1.4)
$$
where $Y$ stands for the hypercharge of (charged) fermions.
In turn this lagrangean may arise as a low-energy part of the Higgs field 
model or of a theory with dynamically generated fermion masses. 
After fermion quantization the corresponding effective action
at low momenta  or for  heavy fermions yields the pertinent 
Wess-Zumino vertex as a main contribution.
 
It can be shown as well - see the Appendix - that, in the limit
of small gauge particle momenta, quadratic kinetic terms for the fields
$\chi$ and $\tilde\chi$ are also generated by quantum effects.
To sum up, the low momenta regime of the abelian chiral gauge theory (1.1) or
QED with the "chiral mass"term (1.4) is 
faithfully described by the following non-renormalizable effective lagrangean
density: namely,
$$
{\cal L}_{{\rm eff}}[\theta ,F_{\mu\nu}] =
-{1\over 4}F_{\mu\nu}F^{\mu\nu}
+{\kappa\over 2M}\theta\tilde F_{\mu\nu}F^{\mu\nu}
+{1\over 2}\partial_\mu\theta\partial^\mu\theta\ ,
\eqno(1.5)
$$
in which we have set $\theta\equiv M\chi$, standing for a 
pseudoscalar axion-like
field, $M$ is  some reference mass scale,
while $\kappa$ is the dimensionless WZ coupling parameter of the order
$\alpha\sqrt\alpha$ in the case of the chiral gauge model (1.1)
or $\alpha Y$ for QED with chiral mass term, 
$\alpha$ being the fine structure constant.

This latter model may also have a different origin, the pseudoscalar
field being a scalar gravitational [14] or quintessence field [15]
or even associated to the torsion field of a particular [16], divergenceless
type: $T_{\mu\nu\sigma} = \epsilon_{\mu\nu\sigma\rho} \partial^\rho \chi(x);\
\partial^\mu T_{\mu\nu\sigma} = 0$.

Phenomenologically the overall inverse 
coupling of pseudoscalar particles to photons
is actually constrained from laboratory experiments, as well as from
astrophysical and cosmological observations [17] to be more than
$10^{12}\ {\rm GeV}$: namely, we can reasonably suppose our
reference mass $M$ to be of the same very large order of 
magnitude when we remain within the perturbation approach. 

On the other hand, one of the aims of the present paper 
is to show that
 the effective lagrangean (1.5),
which describes quantum effects of the abelian anomalous gauge theory
or QED with chiral mass interaction, at
small momenta $p$ such that $(p/M)\ll 1$, can lead to the
dynamical breaking of the Lorentz symmetry, the non-perturbative phenomenon
which changes drastically the photon spectrum and induces the birefringence
of photons with opposite helicities. In this regime the pseudoscalar
field loses time derivatives in the kinetic term 
(when treated in the static frame) and therefore cannot
describe a propagating particle, thereby making the bounds from [17]
inapplicable.  

The paper is organized as follows. In Section 2 the renormalizable version
of the abelian AWZ model is implemented with the help of higher
derivatives in the kinetic term for photons. The remaining divergences
are analyzed, the $\beta$ function and anomalous dimensions are exactly
calculated. It is proven within the perturbation theory 
that the ghost-like vector-field 
degrees of freedom decouple at small momenta and 
the model is infrared stable and non-trivial at low energies.

In Section 3 the one-photon-loop effective potential for axion field 
is derived in the renormalizable AWZ model by employing the 
$\zeta$-function technique [18],[19]. This effective potential is
shown to possess a minimum at non-zero values of $\partial_\mu \theta$
for large values of normalization scale, {\it i.e.} in the
strong coupling regime. 

This phenomenon of axion field condensation
leads to the Lorentz symmetry breaking, whose consequences for the
photon spectrum are examined in Section 4. In particular, it is 
elucidated that
the tachyon modes appear in the photon spectrum [2] and photons of 
different helicities propagate with different phase velocities which leads
to the birefringence of arbitrarily polarized photon waves.
The instability of the Fock vacuum arises if v.e.v. of 
$\partial_\mu \theta$ is a time-like vector, whereas for the space-like
one the consistent LSB may be induced by infrared radiative effects.

In our Conclusion the perturbation theory in the symmetry broken phase
is shortly outlined and the propagators for distorted photons is obtained.

\bigskip
\noindent
{\bf 2.\ Renormalizable Axion-Wess-Zumino model.}
\medskip

The renormalizable abelian vector-field model 
(in the euclidean space) we consider is
described  by the lagrangean density which contains the 
Wess-Zumino coupling of pseudoscalar axion and abelian gauge field,
as well as higher derivative kinetic term for the abelian gauge field:
$$
\eqalign{
{\cal L}_{AWZ}\ & =\ {1\over 4M^2} \partial_{\rho}F_{\mu\nu}
\partial_{\rho}F_{\mu\nu}\  +\
{1\over 4} F_{\mu\nu} F_{\mu\nu}\ 
\ + \ {1\over 2\xi} (\partial_{\mu}A_{\mu})^2\cr
& + {1\over 2}\partial_{\mu}\theta \partial_{\mu}\theta\ -\
i{\kappa\over 2M} \theta
F_{\mu\nu} \tilde F_{\mu\nu}\ ,  \cr}
\eqno(2.1)
$$
where $\tilde F_{\mu\nu}
 \equiv (1/2)\epsilon_{\mu\nu\rho\sigma} F_{\rho\sigma}$,
some suitable  dimensional scale $M$ is introduced,
 $\kappa$ and $\xi$ being the dimensionless coupling and gauge fixing 
parameters respectively. 

The Wess-Zumino interaction  can
be equivalently represented in the following form,
$$
\int d^4x\ {\kappa\over 2M}\ \theta\ F_{\mu\nu} \tilde F_{\mu\nu} \ =\
- \ \int d^4x\,{\kappa\over M}\, \partial_{\mu}\theta\, A_{\nu}
\tilde F_{\mu\nu}\ ,
\eqno(2.2)
$$
at the level of the classical action. Therefore the pseudoscalar field
is involved into the dynamics only through its gradient
$\partial_{\mu}\theta(x)$ due to topological triviality of abelian
vector fields.

From the above lagrangean it is easy to derive the Feynman rules: namely,
the free vector field propagator reads

$$
D_{\mu\nu}(p)=-M^2{d_{\mu\nu}(p)\over p^2(p^2+M^2)}+{\xi\over p^2}
{p_\mu p_\nu\over p^2}\ ,
\eqno(2.3)
$$
with $d_{\mu\nu}(p)\equiv -\delta_{\mu\nu}+(p_\mu p_\nu/p^2)$ being the 
transversal projector; the free axion propagator is the usual 
$D(p)=(p^2)^{-1}$ and the axion-vector-vector WZ vertex 
turns out to be given by
$$
V_{\mu\nu}(p,q,r) = - i(\kappa/M)\epsilon_{\mu\nu\rho\sigma}p_\rho q_\sigma\ ,
\qquad (p+q+r=0)
\eqno(2.4)
$$
all momenta being incoming, $r$ being axion's momentum.
It is worthwhile to recall that the Fock space of asymptotic states, in the
minkowskian formulation of
the present model, exhibits an indefinite metric structure. Actually,
from the algebraic identity
$$
{M^2\over p^2(p^2+M^2)}\ \equiv\ {1\over p^2}\ -\ {1\over p^2+M^2}\ ,
$$
it appears that negative norm states are generated by the 
asymptotic vector field transversal
component   with ghost-mass $M$; in addition, the longitudinal component 
of the
asymptotic vector field also gives rise to negative norm states.
 
Now let us develop the power counting analysis of the superficial degree of 
divergence within the model. The number of loops is as usual $L= I_v+I_s-V+1$,
$I_{v(s)}$ being the number of vector (scalar) internal lines and $V$ the number
of vertices. Next we have $2V=2I_v+E_v$ and $V=2I_s+E_s$, where $E_{v(s)}$ is
the number of vector (scalar) external lines. As a consequence the overall
UV behaviour of a graph $G$ is provided by
$$
\omega (G)=\, 4L-4I_v-2I_s+2V-E_s-E_v\,=\,4-2E_v-E_s-2I_v+2I_s\ ,
\eqno(2.5)
$$
and therefrom we see that the {\sl only} divergent graph corresponds to
$I_s=1,\,I_v=1,\,E_s=0,\,E_v=2$ and it turns out to be the one
loop vector self-energy\footnote{$^1$}
{Actually, the tadpole $E_s=I_v=1,\,I_s=0$ 
indeed vanishes owing to the tensorial 
structure of the AWZ-vertex}. Thus we conclude that the model is 
super-renormalizable. We notice that the number of external vector
lines has to be even. The computation of the divergent self-energy can
be done using dimensional regularization (in $2\omega$ dimensional euclidean 
space) and gives
$$
\Pi^{(1)}_{\mu\nu}(p)\,=\,{g\over 16\pi\epsilon} p^2d_{\mu\nu}(p)
+ \hat\Pi^{(1)}_{\mu\nu}(p)\ ,
\eqno(2.6)
$$
with $\epsilon\equiv 2-\omega,\,g\equiv (\kappa^2/4\pi)$, 
while the finite part reads
$$
\eqalign{
\hat\Pi^{(1)}_{\lambda\nu}(p)\,  = & - {g\over 16\pi}p^2d_{\lambda\nu}(p)
\times\left\{\ln{M^2\over 4\pi\mu^2} - \psi(2)+{2\over 3}\left[
1+{p^2+M^2\over p^2}
\ln\left(1+{p^2\over M^2}\right)\right]\right.\cr
& - {M^2\over 3p^2}\left[1-{p^2+M^2\over p^2}
\ln\left(1+{p^2\over M^2}\right)\right]\cr
& \left.-{p^2\over 3M^2}\left[1- {p^2+M^2\over
p^2}\ln\left(1+{p^2\over M^2}\right)+\ln {p^2\over M^2}\right]\right\}\ .\cr}
\eqno(2.7)
$$
where $\mu$ denotes as usual the mass parameter in the 
dimensional regularization.
It follows therefore that the single countergraph to be added, in order to make 
finite the whole set of proper vertices, is provided by the 2-point 1PI
structure
$$
\Gamma_{\lambda\nu}^{({\rm c.t.})}(p)\equiv -\left.\Pi^{(1)}_{\lambda\nu}(p)
\right|_{{\rm div}}\,=\,-{1\over 16}{g\over \pi}p^2d_{\lambda\nu}(p)
\left[{1\over \epsilon}\,+\,F_1\left(\epsilon,{M^2\over 4\pi\mu^2}\right)
\right]\ ,
\eqno(2.8)
$$
in which $F_1$ denotes the scheme-dependent finite part (when $\epsilon\to 0$)
of the countergraph.

As a result it is clear that we can write the renormalized lagrangian 
in the forms
$$
\eqalign{
{\cal L}_{AWZ}^{({\rm ren})}\ & =\ 
{1\over 4M_0^2} \partial_{\rho}F^{(0)}_{\mu\nu}
\partial_{\rho}F^{(0)}_{\mu\nu}\  +\
{1\over 4} F^{(0)}_{\mu\nu} F^{(0)}_{\mu\nu}\ 
\ + \ {1\over 2\xi_0} (\partial_{\mu}A^{(0)}_{\mu})^2\cr
& + {1\over 2}\partial_{\mu}\theta \partial_{\mu}\theta\ -\
i\mu^\epsilon{\kappa_0\over 2M_0} \theta
F^{(0)}_{\mu\nu} \tilde F^{(0)}_{\mu\nu}\cr
& = {1\over 4M^2} \partial_{\rho}F_{\mu\nu}
\partial_{\rho}F_{\mu\nu}\  +\
{Z\over 4} F_{\mu\nu} F_{\mu\nu}\ 
\ + \ {1\over 2\xi} (\partial_{\mu}A_{\mu})^2\cr
& + {1\over 2}\partial_{\mu}\theta \partial_{\mu}\theta\ -\
i\mu^\epsilon{\kappa\over 2M} \theta
F_{\mu\nu} \tilde F_{\mu\nu}\ ,  \cr}
\eqno(2.9)
$$
where, due to super-renormalizability,
the exact wave function renormalization constant $Z$ is provided by
$$
Z\ =\ c_0\left(g ,{M\over \mu};\epsilon\right)\ +\ 
{1\over \epsilon}c_1(g)\ ;
\eqno(2.10)
$$
here we can write, up to the one loop approximation, 
$$
\eqalign{ 
& c_0\left(g,{M\over \mu};\epsilon\right)\ =\ 1\ -\ {g\over
16\pi}F_1\left(\epsilon,{M^2\over 4\pi\mu^2}\right)\ +\ {\cal O}(g^2)\ ,\cr
& c_1(g)\ =\ -{g\over 16\pi}\ .\cr}
\eqno(2.11)
$$
Moreover the relationships between bare and renormalized quantities
turn out to be the following: namely,
$$
\eqalignno{
& A_\mu^{(0)}\ =\ \sqrt{Z} A_\mu\ , &(2.12a)\cr
& M_0\ =\ \sqrt{Z} M\ , &(2.12b)\cr
& \xi_0\ =\ Z \xi\ , &(2.12c)\cr
& \kappa_0\ =\ {\kappa\over \sqrt{Z}}\ ,
\quad g_0\ =\ {g\over Z}\ . &(2.12.d)\cr}
$$
In particular, from the Laurent expansion of eq.~(2.12d), we can write
$$
\kappa_0\ =\ a_0\left(\kappa,{M\over \mu};\epsilon\right)\ 
+\ {1\over \epsilon} a_1(\kappa)\ ,
\eqno(2.13)
$$
with
$$
\eqalign{
& a_0\left(\kappa,{M\over \mu};\epsilon\right)\ =\ 
\kappa+{\kappa^3\over 128\pi^2}F_1\left(\epsilon,{M^2\over 4\pi\mu^2}\right)
+{\cal O}(\kappa^5)\ ,\cr
& a_1(\kappa)\ =\ {\kappa^3\over 128\pi^2}\ .\cr}
\eqno(2.14)
$$
This entails that, within this model, we can solve the renormalization group
equations (RGE) in the minimal subtraction (MS) scheme $F_1\equiv 0$: namely,
$$
\mu{\partial\kappa\over \partial\mu}\ =\ -\epsilon\kappa-a_1(\kappa)+
\kappa{d\over d\kappa}a_1(\kappa)\ ,
\eqno(2.15)
$$
to get the exact MS prescription $\beta$-function
$$
\beta (\kappa)\ =\ {\kappa^3\over 64\pi^2}\ ,
\eqno(2.16)
$$
telling us, as expected, that $g=0$ is an IR stable fixed point.
It follows that we can integrate eq.~(2.15) and determine the running coupling
exact behaviour
$$
g(\mu)\ =\ {g(\mu_0)\over 1-[g(\mu_0)/8\pi]\ln (\mu/\mu_0)}\ .
\eqno(2.17)
$$
Furthermore, always from eq.s~(2.12a-d) and within the MS prescription, 
it is straightforward to recognize the remaining RG coefficients to be
$$
\eqalignno{
& \gamma_M\ \equiv\ {1\over 2}\mu{\partial\ln M^2\over \partial\mu}\ =\
-{g\over 16\pi}\ , &(2.18a)\cr
& \gamma_d\ \equiv\ {1\over 2}\mu{\partial\ln Z\over \partial\mu}\ =\
{g\over 8\pi}\ , &(2.18b)\cr
& \gamma_\xi\ \equiv\ \mu{\partial\ln\xi\over \partial\mu}\ =\
-{g\over 4\pi}\ . &(2.18c)\cr}
$$
In conclusion, we are able to summarize the asymptotic behaviour
of the ghost-mass parameter $M$ and of the gauge-fixing parameter $\xi$ at
large distances, where perturbation theory is reliable in the model
we are considering and within the MS renormalization scheme. Actually,
if we set $s\equiv (\mu/\mu_0)$, we can easily derive
$$
\eqalignno{
& \bar g(s;g)\ =\ {g\over 1-(g/8\pi)\ln s}\
\buildrel s\to 0 \over \sim\ -{8\pi\over \ln s}\ , &(2.19a)\cr
& \bar M(s;M,g)\ =\ M\sqrt{1-{g\over 8\pi}\ln s}\
\buildrel s\to 0 \over \sim\ M\sqrt{{g|\ln s|\over 8\pi}}\ ,
& (2.19b)\cr
& \bar\xi (s;\xi,g)\ =\ \xi+\ln\left(1-{g\ln s\over 8\pi}\right)\
\buildrel s\to 0 \over \sim\ \xi+2\ln\left({g\over 4\pi}|\ln s|\right)\ 
, &(2.19c)\cr}
$$
showing that longitudinal as well as ghost-like transverse vector
field degrees of freedom decouple at small momenta where perturbation 
theory has to be trusted. Owing to this asymptotic decoupling of negative 
norm states, within the domain of validity of perturbation theory, the present 
super-renormalizable model might be referred to as {\sl asymptotically unitary}.

Now, since eq.~(2.17) holds exactly within the MS renormalization prescription,
it is important to analyze the matter of triviality in the present model.
First of all it is worthwhile to notice, taking eq.s~(2.12b,d) into account,
that the quantity $\kappa_0 M_0=\kappa M\equiv 4\pi M_{{\rm inv}}$ is a 
RG-invariant mass parameter. Furthermore, it is useful to rewrite the
renormalized lagrangean in the form
$$
\eqalign{
{\cal L}_{AWZ}^{({\rm ren})} 
& = {1\over 4M^2} \partial_{\rho}F_{\mu\nu}
\partial_{\rho}F_{\mu\nu}\  +\
{Z\over 4} F_{\mu\nu} F_{\mu\nu}\ 
\ + \ {1\over 2\xi} (\partial_{\mu}A_{\mu})^2\cr
& + {1\over 2}\partial_{\mu}\theta \partial_{\mu}\theta\ -\
i\mu^\epsilon{g_0\over 2M_{{\rm inv}}} \theta
F_{\mu\nu}^{(0)} \tilde F_{\mu\nu}^{(0)}\ .  \cr}
\eqno(2.20)
$$
Remembering that in the MS scheme we have the following relationships: namely,
$$
g_0(\epsilon)\ =\ Z^{-1}_{MS} g_{MS}(\mu)\ =\ 
{g_{MS}(\mu)\over 1-[g_{MS}(\mu)/16\pi\epsilon]}\ ,
\eqno(2.21)
$$
where $g_{MS}(\mu)$ is given by eq.~(2.17), we are indeed allowed to 
specify arbitrarily the mass $M_{{\rm inv}}(\epsilon)$, which turns out
to be some {\sl free} mass parameter, analytic when $\epsilon\to 0$, in the
present model.

Now, let us suppose $\epsilon>0,\ g_0 (\epsilon)\ll 1\ (\sim 10^{-2}\
{\it e.g.})$; when $g_{MS}(\mu)=16\pi\epsilon>0$, then eq.~(2.21) can
no longer be fulfilled unless $g_{MS}(\mu)=g_0(\epsilon)=0,\ 
\forall \mu>0$. On the other hand, this situation 
does not entail triviality of the model as we can always choose 
$M_{{\rm inv}}\to 0$ in such a way that the ratio 
$[g_0(\epsilon)/M_{{\rm inv}}(\epsilon)]=
(1/M_*)\not= 0$. As a consequence the non-trivial renormalized model is most
suitably parametrized as follows: namely,
$$
\eqalign{
{\cal L}_{AWZ}^{({\rm ren})} 
& = {\rho\over 4M_*^2} \partial_{\rho}F_{\mu\nu}
\partial_{\rho}F_{\mu\nu}\  +\
{Z(\epsilon)\over 4} F_{\mu\nu} F_{\mu\nu}\ 
\ + \ {1\over 2\xi} (\partial_{\mu}A_{\mu})^2\cr
& + {1\over 2}\partial_{\mu}\theta \partial_{\mu}\theta\ -\
\mu^\epsilon{i\over 2M_*} \theta
F_{\mu\nu} \tilde F_{\mu\nu}\ ,  \cr}
\eqno(2.22)
$$
in terms of the free RG-invariant mass $M_*$ and of the unitarity violation
running parameter $\rho\equiv (M_*^2/M^2)$, 
which asymptotically vanishes at large
distances as already stressed. We notice that it is just the above 
RG-invariant free mass that has to be eventually identified with the
"physical value" $M_*\ge 10^{12}$ GeV, as discussed in Ref.~[17].
However, it is important to gather that what has been discussed in the present
section is actually pertinent to the unbroken Lorentz covariant phase.
As a matter of fact, we shall see in the next section that quantum radiative
effects may lead, in the present model, to the onset of another phase
in which Lorentz symmetry appears to be dynamically broken and a non-trivial
v.e.v. for the quantity $\partial_\mu \theta$ arises.
\bigskip
\noindent
{\bf 3.\ Effective potential.}
\medskip

We are ready now to investigate a further interesting feature of this
simple but non trivial model: the occurrence of the spontaneous breaking
at the quantum level
of the $SO(4)$-symmetry in the euclidean version, or the 
$O(3,1)^{++}$ space-time symmetry in the minkowskian case.
As a matter of fact, we shall see in the following that the effective
potential for the pseudoscalar axion field $\theta$ may exhibit 
nontrivial minima and, consequently, some privileged direction has
to be fixed by boundary conditions, in order to specify 
the true vacuum of the model.
More interesting, those non trivial minima lie within the perturbative domain.
Since we are looking for the effective potential of the pseudoscalar field,
we are allowed to ignore the renormalization constant $Z(\epsilon)$ in
eq.~(2.22) and restart
from the classical action in four dimensions.

 The axion
background field generating functional is defined as
$$\eqalign{
&{\cal Z}[\theta]\ \equiv\ {\cal N}^{-1}\int [{\cal D}A_\mu]\
\exp\left\{-{\cal A}_{AWZ}[A_\mu,\theta]\right\}; \cr
&{\cal A}_{AWZ}[A_\mu,\theta] \equiv \int d^4x
\left({\cal L}_{AWZ} - A_\mu J_\mu \right),\cr}
\eqno(3.1)
$$
where we have included the photon coupling to (external) matter sources
$J_\mu$.
The classical field configurations $\bar A_\mu (x)$ are the solutions of the
Euler-Lagrange equations
$$
{\delta {\cal A}_{AWZ}[A_\mu,\theta]\over \delta A_\mu (x)}\ =\
K_{\mu\nu}[\theta]\bar A_\nu (x)\ =\  J_\mu\ ,
\eqno(3.2)
$$
with ($\triangle\equiv \partial_\mu\partial_\mu$)
$$
\eqalign{
K_{\mu\nu}[\theta]\ &\equiv\ 
\left(\rho{\triangle\over M_*^2}-1\right)
\left(\delta_{\mu\nu}\triangle-\partial_\mu\partial_\nu\right)\cr
& -\ {1\over \xi}\partial_\mu\partial_\nu - {2\over M_*}
\epsilon_{\lambda\mu\sigma\nu}\partial_\lambda\theta (x)
(-i\partial_\sigma)\ ,\cr}
\eqno(3.3)
$$
being an elliptic invertible local differential operator. 
After integrating over photon fluctuations $A_\mu(x) - \bar A_\mu(x)$
we eventually obtain
$$
{\cal Z}[\theta]\ \equiv\ {\cal N}^{-1}
\exp\left\{-{\cal A}_{AWZ}[\bar A_\mu,\theta]\right\}\times
\left({\tt det}\parallel {\cal K}_{\mu\nu}[\theta]\parallel\right)^{-1/2}\ ,
\eqno(3.4)
$$
with ${\cal N}={\cal Z}[\theta=0]$ and where the dimensionless operator has been
introduced: namely,
$$
\eqalign{
{\cal K}_{\kappa\nu}[\theta]\ &\equiv\ \mu^{-2}K_{\kappa\nu}[\theta]\cr
=\ &\top_{\kappa\nu}{\triangle\over \mu^2}
\left(\rho{\triangle\over M_*^2}-1\right)
-{1\over \xi}{\triangle\over \mu^2}\ell_{\kappa\nu}\cr
&\ -{2\over \mu^2}\epsilon_{\kappa\nu\lambda\sigma}\eta_\lambda (x)
(-i\partial_\sigma)\ ,\cr}
\eqno(3.5)
$$
where we have set
$$
\eqalignno{
& \top_{\mu\nu}\ \equiv\ \delta_{\mu\nu}\ -\ {\partial_\mu\partial_\nu\over 
\triangle}\ , &(3.6a)\cr
& \ell_{\mu\nu}\ \equiv\ {\partial_\mu\partial_\nu\over \triangle}\ ,
&(3.6b)\cr
& \eta_\mu (x)\ \equiv\ {\partial_\mu\theta (x)\over M_*}\ , &(3.7)\cr}
$$
in which $\mu$ represents the subtraction point, $i.e.$ the momentum scale
at which the effective action is defined, whose actual value is constrained
by physical requirements as we shall see below.

We want to evaluate the determinant into eq.~(3.4) for  constant 
vector $\eta_\mu$; to this aim we can rewrite the relevant
operator into the form
$$
{\cal K}_{\kappa\nu}(\eta)\ \equiv\ 
{\triangle\over \mu^2}
\left\{ - \top_{\kappa\nu}\left(1-\rho{\triangle\over M_*^2}\right) 
- {1\over \xi}\ell_{\kappa\nu}\right\}
+ {{\cal E}_{\kappa\nu}(\eta)\over \mu^2} ,
\eqno(3.8)
$$
with
$$
{\cal E}_{\mu\nu}(\eta)\ \equiv\ 
- 2\epsilon_{\mu\nu\lambda\sigma}\eta_\lambda
(-i\partial_\sigma)\ .\eqno(3.9)
$$
From the conjugation property
$$
\left({\cal E}^\dagger\right)_{\mu\nu}\ =\ -{\cal E}_{\mu\nu}\ ,
\eqno(3.10)
$$
it follows that
$$
\left({\cal K}^\dagger[\eta]\right)_{\mu\nu}\ =\ 
\left({\cal K}[-\eta]\right)_{\mu\nu}\ ,
\eqno(3.11)
$$
which shows that the the relevant operator is {\sl normal}.
As a consequence, after compactification of the
euclidean space, we can safely define its complex power [18] and its
determinant [19] by means of the $\zeta$-function technique:
namely,
$$
\eqalign{
{\tt det}\left\Vert{\cal K}[\eta]\right\Vert\ &=\ 
\left( {\tt det}\left\Vert{\cal K}[\eta]{\cal K}^\dagger 
[\eta]\right\Vert\right)^{1/2}\cr
&\equiv\ \left.\exp\left\{-{1\over 2}{d\over ds}\zeta_H (s;\eta)
\right\}\right\vert_{s=0}\ ,\cr}
\eqno(3.12)
$$
where we have set\footnote{$^2$}{The same regularized determinant is
obtained by considering $H^\prime[\eta]\equiv {\cal K}^\dagger[\eta]{\cal K} 
[\eta]$.}
$$
\eqalignno{
\left(H[\eta]\right)_{\mu\nu}\ &\equiv\ 
\left({\cal K}[\eta]\right)_{\mu\lambda}
\left({\cal K}^\dagger [\eta]\right)_{\lambda\nu}\ ,&(3.13)\cr
\zeta_H (s;\eta)\ &\equiv\ {\tt Tr}\left(H[\eta]\right)^{-s}\ .&(3.14)\cr}
$$
Going into the momentum representation, it is easy to obtain from eq.~(3.13)
the Fourier transform of our relevant operator: namely,
$$
\left(\tilde H[p;\eta]\right)_{\mu\nu}=\left({p^2\over \mu^2}\right)^2
\left\{- \left(1+\rho{p^2\over M_*^2}\right)^2 
t_{\mu\nu}+{1\over \xi^2}l_{\mu\nu}
\right\}- 4{(\eta\cdot p)^2-\eta^2 p^2\over \mu^4} e_{\mu\nu}\ ,
\eqno(3.15)
$$
in terms of the projectors
$$
\eqalignno{
t_{\mu\nu} & =\delta_{\mu\nu}-{p_\mu p_\nu\over p^2}\ ,&(3.16a)\cr
l_{\mu\nu} & ={p_\mu p_\nu\over p^2}\ ,&(3.16b)\cr
e_{\mu\nu} & \equiv \left\{{\rm {\bf e}}_2(p;\eta)\right\}_{\mu\nu}\cr
& ={p^2\over (\eta\cdot p)^2-\eta^2 p^2}
\left\{- \eta^2 t_{\mu\nu}+\eta_\mu\eta_\nu+
{(\eta\cdot p)^2\over p^2}\delta_{\mu\nu}-{\eta\cdot p\over p^2}
(\eta_\mu p_\nu+ \eta_\nu p_\mu)\right\}\ ;&(3.16c)\cr}
$$
notice that the following properties hold
$$
e_{\mu\nu}p_\nu=0\,\quad e_{\mu\nu}\eta_\nu=0\ ,\quad e_{\mu\nu}t_{\nu\lambda}=
e_{\mu\lambda}\ .
\eqno(3.17)
$$
Taking all those definitions and properties carefully into account, it is 
straightforward to rewrite the relevant operator according to the 
orthogonal decomposition as follows
$$
\tilde H[p;\eta]=\tilde H_0(p)\left\{{\rm {\bf I}d}_4-{\rm {\bf e}}_2
(p;\eta)+{\rm {\bf e}}_2 (p;\eta)
\tilde R[p;\eta]\right\}\ ,
\eqno(3.18)
$$
in which
$$
\eqalignno{
\left(\tilde H_0(p)\right)_{\mu\nu}&=\left({p^2\over \mu^2}\right)^2
\left\{-\left(1+\rho{p^2\over M_*^2}\right)^2 t_{\mu\nu}+{1\over \xi^2}l_{\mu\nu}
\right\}\ ,&(3.19)\cr
\tilde R[p;\eta]&=\left(1+
{4\left[(\eta\cdot p)^2-\eta^2 p^2\right]\over 
(p^2)^2 [1+\rho (p^2/M_*^2)]^2}\right)\ ,&(3.20)\cr}
$$
while the projector ${\rm {\bf e}}_2(p;\eta)$ onto a two-dimensional
subspace fulfils
$$
{\tt tr}{\rm {\bf e}}_2(p;\eta)\ =\ 2\ ,\quad 
{\rm {\bf e}}_2(p;\eta=0)\ =\ 0\ , 
\eqno(3.21)
$$
where ${\tt tr}$ means contraction over four-vector indices.

As a consequence, from eq.s~(3.4) and (3.12), we can write eventually
$$
\eqalign{
{\cal Z}[\eta_\mu]&=\exp\left\{-{\cal A}_{AWZ}[\bar A_,\eta]+
{\cal A}_{AWZ}[\bar A_,\eta=0]\right\}\cr
&\times\left\{{{\tt det}\Vert
H_0\left({\rm {\bf I}d}_4-{\rm {\bf e}}_2+{\rm {\bf e}}_2 R (\eta)
\right)\Vert\over
{\tt det}\Vert H_0\Vert}\right\}^{-1/4}\ ;\cr}
\eqno(3.22)
$$
here $H_0$ and $R(\eta)$ stand, obviously, for the integro-differential
operators whose Fourier transforms are given by eq.s~(3.19a-b)
respectively.

We can definitely obtain  
$$
\eqalign{
{\cal W}[\eta_\mu ,\rho]=-\ln{\cal Z}[\eta_\mu,\rho]&\equiv
{\cal A}_{AWZ}[\bar A_,\eta,\rho]-
{\cal A}_{AWZ}[\bar A_,\eta=\rho=0]\cr
&-{1\over 4}{d\over ds}\zeta_h (s=0;\eta,\rho)+
{1\over 4}{d\over ds}\zeta_{h_0} (s=0)\ ,\cr}
\eqno(3.23)
$$
in which
$$
\zeta_h (s;\eta,\rho)=2({\rm vol})_4\mu^{4s}\int {d^4 p\over (2\pi)^4}
\left\{(p^2)^2\left(1+\rho{p^2\over M_*^2}\right)^2
+ 4\left((\eta\cdot p)^2-\eta^2 p^2\right)\right\}^{-s}\ ,
\eqno(3.24)
$$
while, obviously, $\zeta_{h_0} (s)=\zeta_h (s;\eta=\rho=0)$.
The effective potential for constant $\eta_\mu$ appears
eventually to be expressed as
$$
{\cal V}_{{\rm eff}}(\eta,\rho)\ \equiv\  {1\over 2} M_*^2\eta^2 -
{1\over ({\rm vol})_4}\left\{
{1\over 4}{d\over ds}\zeta_h (s=0;\eta,\rho)-
{1\over 4}{d\over ds}\zeta_{h_0} (s=0)\right\}\ ,
\eqno(3.25)
$$
and therefore we have to carefully compute the integral in eq.~(3.24).
To this aim, it is convenient to select a coordinate system in which
$$
p_\mu\ =\ ({\bf p},p_4)\ ,\quad p_4\ =\ {\eta\cdot p\over 
\sqrt{\eta^2}}\ ,
\eqno(3.26)
$$
in such a way that, after rescaling variables as
${\bf x}=({\bf p}/\mu),\ y=(p_4/\mu)$, we obtain
$$
\eqalign{
&\zeta_h (s;\eta,\rho)\ 
=\ {4\mu^4({\rm vol})_4\over (2\pi)^4\Gamma (s)}\times\cr
&\int_0^\infty d\tau\ \tau^{s-1}\int_0^\infty dy\int d^3x\
\exp\left\{-\tau\left({\bf x}^2+y^2\right)^2\left(1+\varrho\left({\bf x}^2+
y^2\right)\right)^2+\tau\upsilon^2{\bf x}^2\right\}\ ,\cr}
\eqno(3.27)
$$
where $\varrho\equiv \rho(\mu/M_*)^2$ and 
$\upsilon_\nu\equiv (2/\mu)\eta_\nu$.
A straightforward calculation leads eventually to the following integral
representation \footnote{$^3$}{We notice that, from the integral 
representation (3.28) for ${\rm Re}s<1$, it turns out that $\zeta_{h_0}(s)$ is
regularized to zero.} 
[20]: namely,
$$
[\mu^4({\rm vol})_4]^{-1}\zeta_h (s;\eta,\rho)={(\upsilon^2)^{2-2s}\over 8\pi^2}
\int_0^\infty dt\ {t^{1-2s}\over \left(1-\varrho\upsilon^2 t\right)^{2s}}\
_2F_1\left({3\over 2},s;2;{-1\over t\left(1-\varrho\upsilon^2 t\right)^2}\right)\ .
\eqno(3.28)
$$

Let us first analyze the case $\rho=0$, which corresponds to the low-energy
unitary regime; in this limit, the integration in the previous formula can be
performed explicitly ($1<{\rm Re}s<(7/4)$) to yield
$$
[\mu^4({\rm vol})_4]^{-1}\zeta_h (s;\eta,\rho=0)=
{(\upsilon^2)^{2-2s}\over 16\pi^2\sqrt\pi}{2^{4s-4}\over (s-1)}
{\Gamma[s-(1/2)]\Gamma[(7/2)-2s]\over \Gamma[(5/2)-s]}\ .
\eqno(3.29)
$$
In the present case $\rho\to 0$, the effective potential for constant 
$\eta_\mu$ within the $\zeta$-function regularization is given by
$$
\eqalign{
 {\cal V}_{{\rm eff}}(\eta,\rho=0) & = {1\over 2} M_*^2 \eta^2
-{1\over ({\rm vol})_4}
{1\over 4}{d\over ds}\zeta_h (s=0;\eta,\rho=0)\cr
& =
{5\mu^4\over 32\pi^2}\left\{az + z^2\left(\ln z +{7\over 30}\right)\right\}
\ ,\cr}
\eqno(3.30)
$$
where $a\equiv (16\pi^2M_*^2/5\mu^2)$
and $z\equiv (\upsilon_\nu\upsilon_\nu/4)=
\eta_\nu\eta_\nu / \mu^2$. 
We can easily check that the stable $O(4)$-degenerate
non-trivial minima appear for $a\le a_{{\rm cr}}=\exp\{-(37/30)\}
\simeq 0.2913$.
Notice that the latter interval of values of $a$ just corresponds to
$\mu\ge 10.4 M_*$.  

It follows thereof that, for $0\le a\le a_{{\rm cr}}$, the
corresponding symmetry breaking values fulfil
$$
z_0\ge z_{{\rm SB}}\ge z_{{\rm cr}}\ ,\quad
z_{0}\ =\ \exp\{-{11\over 15}\}\simeq 0.480\ ,\quad
z_{{\rm cr}}=a\ .
\eqno(3.31)
$$
We remark that the above result, within the $\zeta$-function regularization,
actually reproduces our previous calculation [4] using
large momenta cutoff regularization. To be more precise, eq.~(3.30) indeed
corresponds to a specific choice of the subtraction terms in the large momenta
cutoff method, something we could call {\sl minimal subtraction for the
effective potential}.\footnote{$^4$}{We recall
that, in general, the $\zeta$-regularized functional determinant 
of elliptic invertible normal operators is defined
up to local polynomials of the background fields.}

It is eventually very interesting to study the dependence 
of the symmetry breaking value $z_{SB}$ upon the parameter
$\rho$, which measures the departure of the model from unitarity.
To this aim, it is necessary to come back to the general expression of eq.~
(3.28) and to make use of the Mellin-Barnes transform for the confluent
hypergeometric function [20]. The result eventually reads
$$
\eqalign{
& [\mu^4 ({\rm vol}_4)]^{-1}\zeta_h (s;\eta,\rho)= {1\over 4\pi^2\sqrt\pi}
{(\upsilon^2)^{2-2s}\over \Gamma (s)}\times\cr
& \left\{\sum_{n=0}^\infty{(-1)^n\over n!}(\varrho\upsilon^2)^{n+2s-2}
{\Gamma (s+n)\Gamma\left(n+{3\over 2}\right)\Gamma (2-2s-n)\Gamma (4s-2+3n)
\over \Gamma (2+n)\Gamma (2s+2n)}\right. +\cr
& \left.\sum_{n=0}^\infty{(-1)^n\over n!}(\varrho\upsilon^2)^{n}
{\Gamma (2-s+n)\Gamma\left(n-2s+{7\over 2}\right)\Gamma (2s-2-n)
\Gamma (4-2s+3n)
\over \Gamma (4-2s+n)\Gamma (4-2s+2n)}\right\}\ ,\cr}
\eqno(3.32)
$$
namely, a convergent power series for $\varrho\upsilon^2<(4/27)$, with
$(1/2)<{\rm Re}s<(7/4)$. As a check, we notice that, when $1<{\rm Re}s<(7/4)$,
it is possible to set $\rho\rightarrow 0$ in the previous formula: in
so doing eq.~(3.29) is indeed recovered.

It would be possible, now, to study the behavior of 
${\cal V}_{{\rm eff}}(\eta,\rho)$ up to any order in $\rho$. Nonetheless,
a first indication on the shift of the true minima, in the renormalizable
non-unitary model, is clearly given already at the first order. It reads:
$$
 {\cal V}_{{\rm eff}}(\eta,\rho) =
{5\mu^4\over 32\pi^2}\left\{az + z^2\left(\ln z +{7\over 30}
+ 14\varrho z\ln z +{74\over 15}
\varrho z\right) +
{\cal O}(\varrho^2)\right\}\ .
\eqno(3.33)
$$
In the present case non-trivial minima appear for 
$$
a\le a_{{\rm cr}}(\varrho)=a_{{\rm cr}}(0)+{37\over 3}\varrho
a_{{\rm cr}}^2(0)+\ldots\simeq 0.2913 + 1.046\varrho \ ,
\eqno(3.34)
$$
whose corresponding values are between 
$$
\eqalign{
& z_0(\varrho)\ge z_{{\rm SB}}(\varrho)\ge
z_{{\rm cr}}(\varrho)\ ,\cr
& z_0(\varrho)=\exp\left\{-{11\over 15}\right\}+\varrho
\exp\left\{-{22\over 15}\right\}+ {\cal O}(\varrho^2)\simeq
0.480+0.231\ \varrho\ ,\cr
& z_{{\rm cr}}(\varrho)=a_{{\rm cr}}(0)+{32\over 3}\varrho
a_{{\rm cr}}^2(0)+\ldots\simeq 0.2913 + 0.905\varrho \ . \cr}
\eqno(3.35)
$$

It appears therefore that, within the renormalizable but non-unitary regime,
the dynamical breaking of the $O(4)$ symmetry is enhanced with respect to the
unitary limit $\rho\rightarrow 0$. The persistence of a non-vanishing v.e.v.
of the operator $\partial_\mu\theta$ for any $\rho$ is a quite unexpected
result and, thereby, indeed remarkable.
As a matter of fact, the renormalizable and/or unitary formulations have,
in general, radically different behaviors [6],~[10]. The possible
 occurrence of the
dynamical symmetry breaking for any non-vanishing $\rho$ (renormalizable
model), which remains there in the limit $\rho\rightarrow 0$ (unitary model),
actually unravels that this feature has a deep meaning 
closely connected to infrared properties of the Wess-Zumino interaction
to massless photons, {\it i.e.} to the presence
of the chiral local $U(1)$-anomaly.

\bigskip
\noindent
{\bf 4.\ Lorentz symmetry breaking
in QED due to  CPT-odd interaction}
\medskip
In Electrodynamics,
when one retains its fundamental
character provided by the renormalizability, it is
conceivable to have 
LSB in the
 {\it 3+1} dimensional Minkowski space-time by
the (C)PT-odd Chern-Simons (CS) coupling of photons to
the vacuum [2] mediated
by a constant CS vector $\eta_\mu$ (Carroll-Field-Jackiw model)
\footnote{$^5$}{We notice that our constant vector $\eta_\mu$
is denoted as $s_\mu$ in Ref.~[2]}:
$$
{\cal L}_{{\rm LSB}} =
-{1\over 4}F_{\mu\nu}F^{\mu\nu} 
+ {1\over 2}\epsilon^{\mu\nu\lambda\sigma} A_\mu F_{\nu\lambda}  \eta_\sigma ,
\eqno(4.1)
$$
One can guess that  the CS vector $\eta_\mu$ originates from v.e.v. of
the gradient of axion field $\theta$ in the AWZ model
(2.22) : $\left<\partial_\mu \theta(x)\right>_0 =  M_* \eta_\mu  $.
  
This supplement to  Electrodynamics does not break the
gauge symmetry of the action but splits the dispersion relations
for different photon helicities [2].
As a consequence the {\it linearly} polarized photons exhibit
the birefringence when they
propagate in the vacuum, {\it i.e.} the
rotation of the polarization direction depending on the distance.

If the vector $\eta_\mu$ is time-like, $\eta^2 > 0$, then
this observable effect is isotropic in the preferred frame
(presumably, the Rest Frame of the Universe 
where the cosmic microwave background
radiation is maximally isotropic), since $\eta_\mu = (\eta_0,0,0,0)$.
However it is essentially anisotropic
for space-like  $\eta^2 < 0 $. The first possibility was thoroughly
examined [2],[22] resulting in the bound:
$|\eta_0| < 10^{-33} eV \simeq 10^{-28} cm^{-1}$. Last year
new compilation of data on polarization rotation of
photons from remote radio galaxies was presented [21] and it was 
argued that the 
space anisotropy with  $\eta_\mu \simeq (0, \vec\eta)$ of order
$|\vec\eta| \sim 10^{-32} eV \simeq 10^{-27} cm^{-1}$ exists.
However the disputes about the confidence level
of their result [23] have make it clear that  
this effect still needs a better confirmation.
                                                                            
We can use now the effective potential derived in the previous
section and conclude that
the time-like pattern for the CS interaction is intrinsically inconsistent
as it is accompanied by the creation of tachyonic photon modes
\footnote{$^6$}{The presence of tachyonic modes in the photon spectrum
was mentioned in [2].}
from the vacuum, {\it i.e.} such a vacuum is unstable under the QED radiative
effects [24]. On the contrary, the space-like anisotropy carrying CS vector
does not generate any vacuum instability
and may be naturally induced [4]
by a Coleman-Weinberg mechanism [5] in any scale invariant scenario where
the CS vector is related to v.e.v. of the gradient of a pseudoscalar
field.

Indeed let us analyze the photon energy spectrum which 
can be derived from
the wave equations on the gauge
potential $A_\mu (p)$ in the momentum
representation:
$$
\left( p^2 g^{\mu\nu} + \left({1 \over \xi} - 1 \right)
p^\mu p^\nu + 2 i \epsilon^{\mu\nu\lambda\sigma} \eta_\lambda p_\sigma
\right) A_\nu = - K^{\mu\nu}[\eta]  A_\nu = 0,
\eqno(4.2)
$$
where $K^{\mu\nu}[\eta]$ is given (in Euclidean notations) by (3.3)
and we put $\rho =0$ focusing ourselves on the infrared part of photon spectrum.

It is evident that the CS interaction changes the
spectrum only in the polarization hyper-plane orthogonal to the momentum
$p_\mu$ and the CS vector $\eta_\nu$. The relevant projector on this plane
is $e_{\mu\nu}  \equiv \left\{{\rm {\bf e}}_2(p; \eta)\right\}_{\mu\nu}$
described in (3.16c).
After employing the notation (3.9a),
$
{\cal E}^{\mu\nu} \equiv 2 i \epsilon^{\mu\nu\lambda\sigma} 
\eta_\lambda p_\sigma,
$
one can prove that\footnote{$^7$}{ In
what follows the matrix product
 is provided by  contraction with  $g_{\mu\nu}$.}
$$
{\rm {\bf e}}_2 =
{{\cal E} \cdot {\cal E} \over {\rm N}};
\qquad {\rm N} \equiv 4 ((\eta\cdot p)^2 - \eta^2 p^2),
\eqno(4.3)
$$
and  $ {\cal E}\cdot {\rm {\bf e}}_2 = {\cal E}$. Respectively,
 one can unravel
the energy spectrum of the wave equation (4.2) in terms of two
polarizations of different helicity:
$$
{\rm {\bf e}}_{L, R} = {1 \over 2} \left({\rm {\bf e}}_2  \pm
{{\cal E} \over \sqrt{{\rm N}}} \right);\qquad
{\cal E} \cdot {\rm {\bf e}}_{L, R} = \pm \sqrt{{\rm N}} {\rm\bf P}_{L, R}.
\eqno(4.4)
$$
Then the dispersion relation can be read out of the equation:
$$
(p^2)^2 + 4 \eta^2 p^2 - 4 (\eta\cdot p)^2 = 0.
\eqno(4.5)
$$
From eq.~(4.5) one obtains the different physical properties depending
on whether $\eta^2$ is time-, light- or space-like.

If $\eta^2 > 0$ one can examine photon properties in the rest frame for
the CS vector  $\eta_\mu = (\eta_0,0,0,0)$. Then the dispersion relation,
$$
(p_0)_\pm^2 = \vec p^2 \pm 2 |\eta_0| |\vec p|,
\eqno(4.6)
$$
shows that the upper type of solutions can be interpreted as
describing massless states because
their energies vanish
 for $\vec p =0$ .
Meanwhile  the lower type of distorted photons 
behave as tachyons [2]
 with a real energy for $|\vec p| >  2 |\eta_0|$
(when their phase
velocity are taken into account) .
There are also static solutions with $p_0 =0 \Leftrightarrow
|\vec p| = 2 |\eta_0|$ and
unstable solutions (tachyons) with
a negative imaginary energy for $|\vec p| <  2 |\eta_0|$.

For light-like  CS vectors, $\eta^2 = 0$, one deals with  conventional photons
of shifted energy-momentum spectra for different polarizations:
$$
(p_0 \pm \eta_0 )^2 =(\vec p \pm \vec\eta )^2 .
\eqno(4.7)
$$

If the CS vector is space-like, $\eta^2 < 0$, the photon spectrum is
more transparent
in the static frame where $\eta_\mu = (0, \vec\eta)$. The corresponding
dispersion relation reads:
$$
\left(p_0\right)_{\pm}^2 =
\vec p^2 + 2\vec\eta^2 \pm 2 \sqrt{|\vec\eta|^4 + (\vec\eta \cdot \vec p)^2}.
\eqno(4.8)
$$
It can be checked that in this case $p_0^2 \geq 0$ for all $\vec p$
and neither static nor unstable tachyonic modes do actually arise.
The upper type of solution
describes the massive particle with a mass $m_+ = 2  |\vec\eta|$
for small space momenta $|\vec p| \ll |\vec\eta|$. The lower type of solutions
represents a massless state as $p_0 \rightarrow 0 $ for $|\vec p|
\rightarrow 0$. It might also exhibit the acausal behavior
when $p_\mu  p^\mu < 0$ but, even
in this case $p_0^2 \geq 0$ for all $\vec p$, so that
the unstable tachyonic modes never arise.

In a general frame, for high momenta $|\vec p|\gg |\vec\eta|$, 
$|p_0|\gg \eta_0$,
one obtains the relation:
$$
|p_0| - |\vec p| \simeq \pm (\eta_0 -|\vec\eta| \cos\varphi ),
\eqno(4.9)
$$
where $\varphi$ is an angle between $\vec\eta$ and $\vec p$.
Hence, for a given photon frequency $p_0$, the
phase shift induced by the difference between  wave vectors
of opposite helicities does not depend upon this frequency. Moreover
the linearly polarized waves - a combination of left- and right-handed ones -
 reveal the birefringence phenomenon of rotation of polarization axis with
the distance [2].

Let us now examine the radiative effects induced by the emission of
distorted photons.
In principle the energy and momentum
conservation allows for pairs
of tachyons to be created from the vacuum due to the CS interaction.
Thereby in any model where the CS vector plays a dynamical role, being related
to the condensate of a matter field, one may
expect that, owing to tachyon pairs creation,
the asymptotic Fock's vacuum state becomes unstable and
transforming towards a true non-perturbative state without
tachyonic photon modes. But if we inherit the causal prescription for
propagating physical waves, then the physical states are assigned to possess
the non-negative energy sign. As a consequence
the tachyon pairs
can be created out of the vacuum only if $\eta^2 > 0$. In particular, the static
waves with $p_0 =0\Leftrightarrow |\vec p| = 2 |\eta_0|$
are well produced to destroy the
vacuum state. On the contrary, for $\eta^2 < 0$ the causal prescription for
the energy sign together with the energy-momentum conservation prevent
the vacuum state from photon pair emission.

Thus the decay process holds when static and unstable
tachyonic modes exist. Let us  clarify this point
with the help of the radiatively induced effective potential (3.25), (3.30)
 for the
variable $\eta_\mu$ treated as an average
value
\footnote{$^8$}{
It may be a mean value over  large
volume for slowly varying classical background field or,
eventually, the v.e.v. for an axion-type field.}
of the gradient of a pseudoscalar
field, $\eta^2 = - z \mu^2$ . In this case the infrared normalization scale 
$\mu = \sqrt{-\eta^2/ z}$ has
to be of the order of $10^{-32} \div 10^{-33} eV$ in such a way 
to fit the Carrol-Field-Jackiw-Nodland-Ralston effect [2],~[21].

One can see from
eq.~(3.30) that:

a) if $\eta^2 > 0, $ there appears an imaginary part for the vacuum energy,
$$
{\rm Im}\ {\cal V}_{{\rm ef\/f}} = - {5 \over 32\pi} (\eta_\mu \eta^\mu)^2,
\eqno(4.10)
$$
which characterizes the rate per unit volume
of tachyon pairs production out of the vacuum state;

b) for  $\eta^2 \leq 0$ the effective potential is real and has a maximum
at  $\eta^2 = 0$,  whereas
the true minima arise at non-zero space-like value $\eta^2 = - \mu^2 z_{SB}$
from (3.31).

We conclude that it is unlikely to have the Lorentz symmetry breaking
by the CPT odd interaction (4.1) by means of a time-like CS vector
preserving the rotational invariance in the $\eta_\mu$ rest frame.
Rather intrinsically, the pseudoscalar matter interacting with photons has a
tendency to condensate along a space-like direction. In turn, as we have seen,
it leads to the photon mass formation. Of course, this effect of a
Coleman-Weinberg type
does not yield any explanation for the  magnitude of the scale $\mu$,
which, however, is  implied to be a physical infrared cutoff
of a cosmological origin.
Therefore its magnitude can be thought to be the inverse of the maximal
photon wave length in the Universe: namely, $\lambda_{{\rm max}}
= 1/\mu \simeq 10^{27}cm$.  

\bigskip
\noindent
{\bf 5.\ Conclusions: sketch of perturbation theory in LSB phase}
\medskip
In the previous Section we have used the quasiclassical, one-photon-loop
approach to argue for the existence of the phase with dynamical LSB.
We remark that this phenomenon can be well realized in the perturbative
low-energy domain provided that  the values of involved free parameters
$\mu,\ M_*$ and $\rho$ are appropriately tuned according to eq.s
(3.31) and (3.35). Thus in this feature the AWZ model is closely
analogous to the second one - Abelian Higgs model - in the original 
Coleman-Weinberg paper [5].

The natural question arises about quantum fluctuations with respect to
the LSB vacuum as well as about higher loop corrections.
In order to reply it one should develop the perturbation theory
in the LSB phase.
It can be built with the help of three basic ingredients:
the photon propagator in the background of constant $\eta_\mu$;
\ the AWZ vertex (2.2), (2.4), which remains unchanged,\ and\ the effective
propagator for the $\theta$ field, which  
should be derived from the second variation of
the one-loop effective action ${\cal W} = - \ln{\cal Z}[\theta]$ 
given by (3.23),
in the vicinity of its LSB minimum. The latter definition implies
that the calculation of photon-loop self-energy diagram is to be supplemented
with a particular subtraction of that part 
which is borrowed by the effective $\theta$ field propagator.

Let us display the structure of distorted
photon and $\theta$ propagators.
The photon propagator can be obtained ( in the limit $\rho = 0$) by setting
$$
\partial_\mu\theta =  M_* \eta_\mu + \partial_\mu \vartheta
\eqno(5.1)
$$
in the lagrangean
(2.22) and subsequent inversion of the photon kinetic operator, namely:
$$
\tilde K_{\mu\nu} = -g_{\mu\nu} p^2 + p_\mu p_\nu \left( 1 - {1 \over \xi}
\right) + i \epsilon_{\mu\nu\rho\sigma} \left(\eta^\rho p^\sigma -
p^\rho \eta^\sigma\right) \eqno(5.2)
$$
in Minkowski space-time.The inversion can be 
easily performed by means of a  decomposition
in terms of a suitable complete set of tensors: namely
$$
\eqalign{
\tilde D_{\mu\nu}(p) &=
i\left(1-\xi\right){p_\mu p_\nu\over \left(p^2+i\epsilon\right)^2}
+{i\over \Delta (p,\eta)}\left\{-g_{\mu\nu}p^2+4\eta^2{p_\mu p_\nu\over p^2+
i\epsilon}\right.\cr
&\left. +4 \eta_\mu \eta_\nu -4{\eta\cdot p\over p^2+i\epsilon}\left(
\eta_\mu p_\nu + p_\mu \eta_\nu\right) 
-2i\epsilon_{\mu\nu\rho\sigma}\eta^\rho p^\sigma
\right\}\ ,\cr}
\eqno(5.3)
$$
where 
$$
\Delta (p,\eta) \equiv \Delta_+(p,\eta)\Delta_-(p,\eta)\ ,
\quad \Delta_\pm = p_0^2 - \vec p^2 -2\vec\eta^2\pm \sqrt{|\vec\eta|^4+\left(
\vec\eta\cdot\vec p\right)^2} +i\epsilon\ ,
\eqno(5.4)
$$
and the causal prescription for two poles is indicated.
Herein, in order to make the pole structure of the above propagator
more transparent, we have referred to the static frame where $\eta_\mu =
\left(0,\vec\eta\right)$, according to eq.~(4.8).

In turn the modified kinetic term for pseudoscalar field at low momenta 
is derived from
the second variation of the effective potential (3.30) in terms of (5.1):
$$
{\cal W}^{(2)} \simeq {1\over 2} \int d^4 x \partial_\mu \vartheta (x)
 {1 \over M_*^2}{\delta^2 {\cal V}_{\rm eff} 
\over \delta\eta_\mu \delta\eta_\nu} 
\partial_\nu \vartheta(x)
 = 
{1\over 2} \int d^4 x 
 {5 \over 4 \pi^2M_*^2}  
\left(\eta^\mu\partial_\mu \vartheta(x)\right)^2. 
\eqno(5.5)
$$
This kinetic term does not correspond 
to a relativistic propagating particle as it
does not contain time derivatives. This, of course, is a consequence
of spontaneous LSB in accordance to the Goldstone theorem.
The related "propagator" takes the following form:
$$
\tilde D(p) = {4i \pi^2M_*^2 \over 5} {1 \over (\eta\cdot p)^2} 
\equiv - {4i \pi^2M_*^2 \over 5} {\partial \over \partial(\eta\cdot p)}
{\rm CPV}\left({1 \over \eta\cdot p}\right) 
\eqno(5.6)
$$
where we adopted (as it customary [25]) 
the Cauchy Principal Value prescription for this
space-like singularity. With this prescription,  the emission of
the $\vartheta$ field will never take place and thereby astrophysical
bounds [17] are no longer applicable. One could guess that
in space-time directions orthogonal to $\eta_\mu$ radiatively induced
higher derivative terms play essential role to restore a particle- or
ghost-like dynamics.

Formally with these propagators we do not change the power counting
of Section 2 for UV divergences and the UV renormalizability  is still
available. But with  the (infrared) $\vartheta$-"propagator" (5.6) 
one anticipates  drastical changes in the $\beta$ function and anomalous
dimensions as both the divergent and finite parts of photon
polarization function is no longer presented by eq.s~(2.6) and (2.7).

We conclude that, in contrast to spontaneous breaking of internal symmetries,
LSB leads to substantial modification of the particle dynamics at low momenta
up to disappearance of those particles which implement the
Goldstone theorem. We postpone a more detailed development of 
the perturbation theory in the LSB phase and the discussion of higher-order
loop effects to the next paper.
 
\vfill\eject

\noindent
{\bf Acknowledgements.}
\medskip We are grateful to A. Bassetto for invaluable suggestions and
to R.Tarrach for stimulating discussions.
This work is supported by Italian grant MURST-quota 40\%;
A.A. is also supported by grants RFFI 96-01-00535, GRACENAS 6-19-97
and  by  Spanish Ministerio de Educaci\'on y Cultura.

\bigskip
\noindent
{\bf Appendix.}
\medskip
In this appendix we compute the fermion chiral determinant in the case of
constant homogeneous gauge potential. In so doing, we shall be able to show
that the low momenta effective action for the pseudoscalar axion (the 
longitudinal component of the gauge potential) exhibits a purely quadratic
kinetic term.

The classical action for a Dirac's fermion, in the Minkowski space-time,
coupled with vector and axial-vector gauge potentials reads
$$
{\cal A}_M =\int dx^0d^3{\bf x}\ \bar\psi\left\{i\gamma^\mu\partial_\mu
-m+e\gamma^\mu\left(V_\mu +\gamma_5 A_\mu\right)\right\}\psi\ .
\eqno(A.1)
$$
To our purpose, it is convenient to take the Weyl representation for the
Dirac's matrices: namely,
$$
\gamma^0 =\left(\matrix{{\bf 0}&{\bf Id}_2\cr {\bf Id}_2&{\bf 0}\cr}\right)\ ;
\quad \gamma^j =\left(\matrix{{\bf 0}&\sigma^j\cr -\sigma^j &
{\bf 0}\cr}\right)\ ;
\eqno(A.2)
$$
where $\sigma^j,\ j=1,2,3$, are the Pauli matrices, in such a way that 
$$
\gamma_5\equiv i\gamma^0\gamma^1\gamma^2\gamma^3\ =\
\left(\matrix{-{\bf Id}_2&{\bf 0}\cr {\bf 0}&{\bf Id}_2\cr}\right)\ .
\eqno(A.3)
$$

The effective action is nothing but - up to the factor $(-i)$ - the
logarithm of the determinant of the vector-axial-vector (VAV) Dirac's
operator. Now, in order to have a well defined expression for such a quantity,
it is necessary to make the transition to the euclidean space, {\it i.e.}
to perform the usual Wick's rotation, which leads to the following
euclidean VAV Dirac's operator: namely,
$$
\left(i{\cal D}_E\right) = i\gamma_\mu\partial_\mu + im
+e\gamma_\mu\left(v_\mu +\gamma_5 a_\mu\right)\ ,
\eqno(A.4)
$$
where $\gamma_j = -i\gamma^j ,\ \gamma_4 = \gamma^0$, $(v_\mu ,\ a_\mu)$
being the euclidean VAV potentials. 
If we perform the analytic continuation $a_\mu\longmapsto i\hat a_\mu$,
then the continued euclidean VAV Dirac's operator (A.4) turns out to be 
elliptic [18],
normal and, if zero modes are absent as we now suppose, invertible.
As a consequence, its determinant is safely defined to be [18],~[19]
$$
{\tt det}\left[i{\cal D}_E\right]\equiv
\left.\exp\left\{-{1\over 2}
\left.{d\over ds}\zeta_{\hat h_E}(s)\right|_{s=0}\right\}
\right|_{\hat a_\mu = -ia_\mu}\ ,
\eqno(A.5)
$$
where
$$
\hat h_E\equiv \left(i\hat{\cal D}_E\right)^\dagger 
\left(i\hat{\cal D}_E\right)\ ,
\eqno(A.6)
$$
with
$$
\left(i\hat{\cal D}_E\right)=
i\gamma_\mu\partial_\mu + im
+e\gamma_\mu\left(v_\mu +i\gamma_5 \hat a_\mu\right)\ .
\eqno(A.7)
$$

Let us compute the above quantity in the case of 
homogeneous VAV potentials. We have, in momentum space,
$$
\hat h_E = \left\{ p^2 + m^2 + e^2\left(v^2+\hat a^2\right)
- 2e p_\mu v_\mu\right\}
{\bf Id}_4 -iep_\mu\hat a_\nu\left\{\gamma_\mu ,\gamma_5\gamma_\nu\right\}\
,\eqno(A.8)
$$
and if we choose $\hat a_\mu =\left(0,0,0,\hat a\right)$, we come to the result
$$
\hat h_E = \left\{ p^2 + m^2 + e^2\left(v^2+\hat a^2\right)
- 2e p_\mu v_\mu\right\}
{\bf Id}_4 - 2ie \hat a\left(
\matrix{ p_j\sigma^j & {\bf 0}\cr {\bf 0} & p_j\sigma^j\cr}\right)\ .
\eqno(A.9)
$$

It is now easy to obtain
$$
\eqalign{
& \zeta_{\hat h_E} (s) = {({\rm vol})_4\mu^4\over \pi^{5/2}\Gamma (s)}\times\cr
& \int_0^\infty dt\ t^{s-3/2}\exp\left\{-t\ {m^2+e^2\hat a^2\over 
\mu^2}\right\}
\int_0^\infty dp\ p^2\exp\{-tp^2\}\cosh\left\{2t{e\hat 
a p\over \mu^2}\right\}\ .\cr}
\eqno(A.10)
$$
If we now come back to the original euclidean axial-vector potential - 
{\it i.e.} $\hat a\rightarrow -ia$ - we easily find
$$
\zeta_{h_E}(s)= {m^4({\rm vol})_4\over 4\pi^2}\exp\left\{-s
\ln\left({m\over \mu}\right)^2\right\}{1\over (s-1)(s-2)}\left\{1-
2(s-2)\left({ea\over m}\right)^2\right\}\ .
\eqno(A.11)
$$

Let us consider the chiral limit $v=\pm a\equiv (m\eta/2e)$; 
we can rewrite the previous formula as
$$
\zeta_{h_E}(s)= {m^4({\rm vol})_4\over 8\pi^2}\exp\left\{-s
\ln\left({m\over \mu}\right)^2\right\}{1\over (s-1)(s-2)}\{2-
(s-2) \eta^2\}\ ,
\eqno(A.12)
$$
from which it is easy to read the chiral effective action we were looking
for: namely,
$$
W_\chi = -\ln{\tt det}\left(i{\cal D}_\chi\right) \equiv
{1\over 2}{d\over ds}\zeta (s=0) = {m^4({\rm vol})_4\over 
(4\pi)^2}\chi (\eta)\ ,
\eqno(A.13)
$$
with
$$
\chi (x) = {3\over 2}-\eta^2-(1+\eta^2)\ln\left({m\over \mu}\right)^2\ .
\eqno(A.14)
$$

First, we notice that the first two terms in the RHS of the last formula 
may be ignored, as the effective action is always defined up to polynomials
of momenta and masses. Second, the effective action - in the case of
homogeneous chiral potential - turns out to contain only quadratic terms 
in the chiral potential. Thereof, we can see that functional integration over 
massive left coupled spinors leads, in the low momenta regime, to 
the effective euclidean kinetic lagrangean
$$
{\cal L}_{\rm kin}(\partial_\nu\theta)={1\over 2}\partial_\nu\theta
\partial_\nu\theta\ ,
\eqno(A.15)
$$
as we claimed in Sect.1, whose constant LSB value is
$$
{\cal L}_{\rm kin}(\eta_\nu)= {m^4\over 2\pi^2}\ln\left({m\over \mu}\right)
z_{SB}\ .
\eqno(A.16)
$$

\vfill\eject
\noindent
{\bf References}
\medskip
\item{[1]}\ H. B. Nielsen, I. Picek : Nucl. Phys. {\bf 211B} (1983) 269.
\item{[2]}\ S. M. Carroll, G. B. Field, R. Jackiw : Phys. Rev.
            {\bf D41} (1990) 1231.
\item{[3]}\ S. Coleman, S. L. Glashow : Phys. Lett. {\bf 405B} (1997) 249.
\item{[4]}\ A. A. Andrianov, R. Soldati : Phys. Rev. {\bf D51} (1995) 5961;
 Proc. 11th Int. Workshop QFTHEP96 (St.Petersburg, 1996)
(MSU Publ., Moscow, 1997) 290, {\tt hep-th/9612156}.
\item{[5]}\ S. Coleman, E. Weinberg : Phys. Rev. {\bf D7} (1973) 1888.
\item{[6]}\ A. A. Andrianov, A. Bassetto, R. Soldati : Phys. 
Rev. Lett. {\bf 63}
            (1989); Phys. Rev. {\bf D44} (1991) 2602; Phys. Rev. {\bf D47} 
            (1993) 4801.
\item{[7]}\ L. D. Faddeev : Phys. Lett. {\bf 145B} (1984) 81;
            L. D. Faddeev, S. L. Shatashvili : Theor. Math. Phys. {\bf 60}
            (1986) 206.
\item{[8]}\ R. Jackiw, R. Rajaraman : Phys. Rev. Lett. {\bf 54} (1985) 1219.
\item{[9]}\ K. Harada, I. Tsutsui : Phys. Lett. {\bf 183B} (1987) 311.
\item{[10]}\ C. Adam : Phys. Rev. {\bf D56} (1997) 5135.
\item{[11]}\ Y. Hosotani : Phys. Lett. {\bf 319B} (1993) 332; 
            DPF Conf. (1994) 1403-1405 ({\tt hep-th/9407188});
            UMN-TH-1308/94 ({\tt hep-th/9408148});
            Phys. Rev. {\bf D51} (1995) 2022; 
            Phys. Lett. {\bf 354B} (1995) 396.
\item{[12]}\ P. Cea : Phys. Rev. {\bf D32} (1985) 2785.
\item{[13]}\ V. P. Gusynin, V. A. Miransky, I. A. Shovkovy : Phys. Rev. Lett.
            {\bf 73} (1994) 3499; NSF-ITP-94-74 ({\tt hep-th/9407168});
            Kiev-ITP-95-02E ({\tt hep-ph/9501304}).
\item{[14]}\ W.-T. Ni : Phys. Rev. Lett. {\bf 38} (1977) 301;
\item{    }\ S. M. Carrol, G. B. Field : Phys. Rev. {\bf D43} (1991) 3789.
\item{[15]}\ S. M. Carroll : {\tt astro-ph/9806099}, preprint 
             NSF-ITP/98-063.
\item{[16]}\ V. De Sabbata, M. Gasperini : Phys. Lett. {\bf 83A}
  (1981) 115;
\item{    }\ A. Dobado, A. L. Maroto : Mod. Phys. Lett. {\bf A12}
(1997) 3003.
\item{[17]}\ E. Mass\'o, R. Toldr\`a : Phys. Rev. {\bf D55} (1997) 7967
             and References therein.
\item{[18]}\ R. T. Seeley : Amer. Math. Soc. Proc. Symp. Pure Math. {\bf 10}
            (1967) 288.
\item{[19]}\ S. W. Hawking : Comm. Math. Phys. {\bf 55} (1977) 133;
\item{    }\ E. Elizalde, S. D. Odintsov, A. Romeo, A. A. Bytsenko, S. Zerbini :
             "{\it Zeta regularization techniques with applications}",
             World Scientific, Singapore (1994).
\item{[20]}\ I. S. Gradshteyn, I. M. Ryzhik : "{\it Table of Integrals
            Series and Products}", Academic Press, San Diego (1979).
\item{[21]}\  B. Nodland and J. P. Ralston : Phys. Rev. Lett. {\bf 78}
             (1997) 3043;
             \quad {\it ibid.} {\bf 79} (1997) 1958.
\item{[22]}\  D. Harari and P. Sikivie : Phys. Lett. {\bf 289B} (1992) 67;  
\item{    }\  S. Mohanty and S. N. Nayak : Phys. Rev. {\bf D48} (1993) 1526;
\item{    }\  A. Cimatti, S. di Serego Alighieri, G. B. Field and
              R. A. E. Fosbury : Astrophys. J. {\bf 422} (1994) 562;
\item{    }\  M. P. Haugan and T. F. Kauffmann : Phys. Rev. {\bf D52} 
              (1995) 3168;
\item{    }\  M. Goldhaber and V. Trimble : J. Astrophys. Astr. {\bf 17} 
                (1996) 17.
\item{[23]}\ D. J. Eisenstein and E. F. Bunn : Phys. Rev. Lett. {\bf 79}
             (1997) 1957;          
\item{    }\ S. M. Carrol and G. B. Field :  Phys. Rev. Lett. {\bf 79}
             (1997) 2934;
\item{    }\  J. P. Leahy : {\tt astro-ph/9704285};
\item{    }\ J. F. C. Wardle, R. A. Perley and M. H. Cohen :  
  Phys. Rev. Lett. {\bf 79} (1997) 1801;
\item{    }\ T. J. Loredo, E. E. Flanagan and I. M. Wasserman :
    Phys. Rev. {\bf D56} (1997) 7057;
\item{    }\  B. Nodland and J. P. Ralston : {\tt astro-ph/9706126,
      astro-ph/9708114};
\item{    }\ P. Jain and J. P. Ralston :  {\tt astro-ph/9803164}.
\item{[24]}\ A. A. Andrianov and R. Soldati : {\tt hep-ph/9804448},
             preprint DFUB-98/8; \ UB-ECM-PF-98/10, to appear in Phys. Lett. B.
\item{[25]}\ A. Bassetto, G. Nardelli, R. Soldati: "{\it Yang-Mills
theories in algebraic non-covariant gauges}", World Scientific,
Singapore (1991).
\vfill\eject\end